\documentclass[reprint, aps, pra, floatfix]{revtex4-1}
\usepackage[dvips]{graphicx}

\usepackage{amsmath}
\usepackage{amssymb}
\usepackage{bm}
\usepackage{color}
\usepackage{subfigure}
\usepackage{notes2bib}
\usepackage[colorlinks, citecolor = black, linkcolor = black, urlcolor = blue]{hyperref}

\newcommand{\bk}{\mathbf{k}}
\newcommand{\bq}{\mathbf{q}}
\newcommand{\nn}{\nonumber}
\newcommand{\x}{\hat{\mathcal{X}}}
\newcommand{\hk}{\mathcal{H}_\mathbf{k}}
\newcommand{\h}{\mathcal{H}}
\newcommand{\cua}{\mathcal{U}}
\newcommand{\cva}{\mathcal{V}}
\newcommand{\cu}{\mathcal{U}_t}
\newcommand{\cv}{\mathcal{V}_t}
\newcommand{\tr}{\mathrm{Tr}}
\newcommand{\im}{\mathrm{Im}}
\newcommand{\va}{|\mathbf{0}\rangle_{\mathbf{k}}}
\newcommand{\z}{Z_{\bk}}
\newcommand{\hp}{\hat{\Psi}}

\begin{document}
\title{The Quantum Dynamics of Two-component Bose-Einstein Condensate: an $Sp(4,R)$ Symmetry Approach}
\author{Chang-Yan Wang}
\email{wang.10183@osu.edu}
\affiliation{Department of Physics, The Ohio State University, Columbus, OH 43210, USA}
\author{Yan He}
\email{heyan$_$ctp@scu.edu.cn}
\affiliation{College of Physics, Sichuan University, Chengdu, Sichuan 610064, China}
\begin{abstract}
% The method of geometrization arises as an important tool in understanding the entanglement of quantum fields and the behavior of the many-body system. The symplectic structure of the boson operators provide a natural way to geometrize the quantum dynamics of the bosonic systems of quadratic Hamiltonians, by recognizing that the time evolution operator corresponds to a real symplectic matrix in $Sp(4,R)$ group. We apply this geometrization scheme to study the quantum dynamics of the spinor Bose-Einstein condensate systems, demonstrating that the quantum dynamics of this system can be represented by trajectories in a six dimensional manifold. It is found that the trajectory is quasi-periodic for coupled bosons. The expectation value of the observables can also be naturally calculated through this approach.
The compact groups such as $SU(n)$ and $SO(n)$ groups have been heavily studied and applied in the study of quantum many body systems. However, the non-compact groups such as the real symplectic groups are less touched. In this paper, it is revealed that the quantum dynamics of two-component Bose-Einstein condensate can be described by a \emph{non-compact} real symplectic group $Sp(4,R)$. With this group, an explicit form of the wavefunction in any time of the evolution can be given, meanwhile, this whole time evolution can be shown to correspond to a trajectory in a six-dimensional manifold. By introducing a polar coordinate, we can visualize this six-dimensional manifold in 2d unit disk and reveal the relation between the behavior of the trajectory in this manifold and the eigenenergies of the Hamiltonian. Furthermore, the time evolution of expectation value of a physical observable such as number operator is proven closely related to the behavior of the trajectory in this manifold.

% Compared to the compact groups such as $SU(n)$ and $SO(n)$, the non-compact groups are less touched in the study of quantum many-body physics. Recently, the non-compact group $SU(1,1)$ is used to study the quantum dynamics of one-component Bose-Einstein condensate. Since not only the single-component BEC, but also the two-component one can be implemented in experiment. Then, one may wonder what is the group that can describe the quantum dynamics of two-component BEC. In this paper, it is shown that such a group is the non-compact real symplectic $Sp(4,R)$. With this group, an explicit form of the wavefunction in any time of the evolution can be given, meanwhile, this whole time evolution can be shown to correspond to a trajectory in a six-dimensional manifold. By introducing a polar coordinate, we can visualize this six-dimensional manifold in 2d unit disk and reveal the relation between the behavior of the trajectory in this manifold and the eigenenergies of the Hamiltonian. Furthermore, the time evolution of expectation value of a physical observable such as number operator is proven closely related to the behavior of the trajectory in this manifold.

\end{abstract}

\maketitle

\section{Introduction}
In cold atomic systems, the Feshbach resonances provide an essential tool to control the interaction between atoms \cite{chin_feshbach_2010}. This tool makes it possible to dynamically change the scattering length in Bose-Einstein condensate (BEC) system \cite{clark_collective_2017,feng_correlations_2019, hu_quantum_2019, fu_density_2018}. These experimental advances necessitate the study of quantum dynamics of BEC from theoretical perspective. Just like the $SU(2)$ group can be used to study the evolution of single spin, the $SU(1,1)$ group, which is a non-compact counterpart of $SU(2)$ group, can be used to study the quantum dynamics of single-component BEC  \cite{chen_manybody_2020,lyu_geometrizing_2020,cheng_manybody_2021}. In Ref.\cite{lyu_geometrizing_2020}, by using the $SU(1,1)$ symmetry, it is shown that the quantum dynamics of a single-component BEC under a Bogoliubov Hamiltonian can be represented by a trajectory in a 2d Poincare disk. 

On the other hand, not only the single-component BEC, but also the two-component one can be implemented in experiment \cite{myatt_production_1997,hall_dynamics_1998,hall_measurements_1998,modugno_two_2002}, which also attracts many theoretical studies \cite{ho_binary_1996,ho_spinor_1998,ohmi_boseeinstein_1998,wang_spinorbit_2010,kawaguchi_spinor_2012}. And it is also possible to tune the interaction between different species of atoms through Feshbach resonances \cite{thalhammer_double_2008}. Hence, we may wonder whether there is a group similar to $SU(1,1)$ that can be used to describe the quantum dynamics of two-component BEC.

In this paper, by recognizing that the boson operators naturally realize the symplectic Lie algebra \cite{perelomov_generalized_1986, hasebe_sp_2020}, we reveal that the quantum dynamics of two-component BEC under a Bogoliubov Hamiltonian can be described by the real symplectic group $Sp(4,R)$, which is a highly nontrivial non-compact group. The $Sp(4,R)$ group is defined as the $4\times 4$ complex matrices set $\{\left( \begin{array}{cc} \cua  & \cva \\ \cva^* & \cua ^* \end{array} \right): \cua \cua ^\dagger - \cva\cva^\dagger = I,\ \cua \cva^T = \cva\cua ^T\}$, where $\cua , \cva$ are $2\times 2$ complex matrices \cite{perelomov_generalized_1986,hasebe_sp_2020}. As a non-compact counterpart of $SU(2)$ group, the $SU(1,1)$ has many properties that are similar to $SU(2)$, which makes it relatively easy to understand. However, on the contrary, the $Sp(4,R)$ group has no compact counterpart, which makes it more difficult to study and needs more sophisticated tools.

By using the property of Bogoliubov transformation, it is proven that the ground state of the Bogoliubov Hamiltonian can be represented by the coherent state of $Sp(4,R)$ group and can be parameterized as a point in a more complicated 6-dimensional space $Sp(4,R)/U(2)$. Even though this space has higher dimensions that the one-component case, by introducing a polar coordinate, we can visualize this space in a 2d unit disk. More importantly, we can show that the quenched quantum dynamics of the system starting from a fully condensed state corresponds to a trajectory in this six-dimensional parameter space. Meanwhile, the explicit form of the wavefunction of the system in any time of the evolution is also derived, which is also a coherent state of the $Sp(4,R)$ group.

In the single-component case, the behavior of the parameter in the Poincare disk is determined by the eigenenergy of the Hamiltonian. For real eigenenergy, the parameter stay within the Poincare disk, for imaginary one, the parameter approaches to the boundary. However, for the two-component case, the eigenenergies display a more complicated nature. We explicitly derive the formula for the eigenenergies, and classify them into four cases. For each case, we reveal the behavior of the trajectory in the six-dimensional parameter space. Finally, our formalism allows us to study the dynamics of a physical observable such as the number operator. We explicitly prove the relation between the expectation value of the number operator and the polar coordinate of the six-dimensional parameter space. As a result, this connects the behavior of the trajectory in this parameter space and the behavior of the physical observable.

The rest of this paper is organized as follows. In section \ref{sec:Ham}, we introduce the model Hamiltonian of two-component BEC. Making use of its $Sp(4,R)$ symmetry, we identify its ground state as the $Sp(4,R)$ coherent state, and establish its correspondence to a point in the six-dimensional space $SP(4,R)/U(2)$, which is dubbed as coherent state parameter (CSP) manifold. In section \ref{sec:dyn}, we reveal that the quantum dynamic evolutions can be understood as $Sp(4,R)$ group actions on the CSP manifold. Different types of quantum dynamical evolution are represented by different types of curves in the CSP manifold. In section \ref{eigen}, the formula for the eigenenergies of the Bogoliubov Hamiltonian is explicitly derived. In section \ref{sec:geo}, the CSP manifold is described by a polar coordinate. The evolutions of expectation of observable operators can be understood by a similar group action method and was discussed in section \ref{sec:obs}.  Finally, we conclude in section \ref{sec:con}.

\section{Hamiltonian and ground state parameter manifold}
\label{sec:Ham}

We consider the Hamiltonian of two-component Bose-Hubbard model with synthetic spin-orbit coupling, which can be written as
\begin{eqnarray}
\hat{H}_{BH}&=&\sum_{ij,\bk} \epsilon_{ij}(\bk) a_{i,\bk}^{\dagger} a_{j,\bk} \nn\\
&&+ \frac{1}{2V}\sum_{i,j,\bk,\bk',\bq} g_{ij} a_{i,\bk-\bq}^\dagger a_{j,\bk'+\bq}^\dagger a_{j,\bk'} a_{i,\bk},
\end{eqnarray}
where $i,j = 1,2$ label the two species of bosons, and $g_{ij}$ are coupling constants, and $V$ is volume. Here, we have chosen $\hbar = 1$.

By taking the Bogoliubov approximation, i.e. substituting $a_{i,\bk=\mathbf{0}}$ with the condensate wavefunction $\psi_i = \sqrt{N_i/V}e^{i\theta_i}$, and keep up to the quadratic terms, we get the Bogoliubov Hamiltonian for the two-component BEC system, which can be expressed as $\hat{H}_\mathrm{eff}= \sum_{\bk\neq \mathbf{0}} \hat{H}_\bk + \mathrm{const.}$
\begin{eqnarray} \label{ham}
  \hat{H}_\bk = \hp_\bk^\dagger \hk \hp_\bk, \ \hk = \left( \begin{array}{cc} \xi(\bk) & \eta \\ \eta^* & \xi^*(\bk) \end{array} \right),
\end{eqnarray}
where $\hp_\bk = (a_{1,\bk}, a_{2,\bk}, a_{1,-\bk}^\dagger, a_{2,-\bk}^\dagger)^T$. Here $\xi(\bk)$ is a $2\times 2$ Hermitian matrix given by $\xi_{ij}(\bk) = \frac{1}{2}[\epsilon_{ij}(\bk) + g_{ij} \psi_i \psi_j^*]$, and $\eta(\bk)$ is a complex symmetric matrix given by $\eta_{ij} = \frac{1}{2} g_{ij} \psi_i\psi_j$. We shall emphasize that validity of this Bogoliubov approximation requires that this two-component BEC is miscible, with the miscibility condition is given by $g_{12} < \sqrt{g_{11} g_{22}}$ \cite{pitaevskii_boseeinstein_2016}. Otherwise, this two-component gas would be in phase separation, where the Bogoliubov approximation is invalid.

Note that in this Bogoliubov Hamiltonian, the terms for different $\bk$ is decoupled. By diagonalizing $\kappa \hk$, where $\kappa = \mathrm{diag}(1, 1, -1, -1)$, we can get the quasiparticle excitations, whose annihilation operators are given by the Bogoliubov transformation $\hat{\alpha}_{i,\bk} = \sum_{j=1}^2 u_{\bk,ij} a_{j,\bk} + v_{\bk,ij} a_{j,-\bk}^\dagger$, where $u_\bk, v_\bk$ are $2\times 2$ matrices. And the Bogoliubov transformation has the property \cite{hasebe_sp_2020,perelomov_generalized_1986}
\begin{eqnarray}\label{bogo}
  u_\bk u_\bk^\dagger - v_\bk v_\bk^\dagger = I,\ u_\bk v_\bk^T = v_\bk u_\bk^T,
\end{eqnarray}
where $I$ is the $2\times 2$ identity matrix. Hence, the ground state of the Hamiltonian can be defined as $\hat{\alpha}_{i,\bk} |G\rangle = 0,\forall\  \bk$ and $\ i = 1,2$.

On the other hand, if we define 
\begin{eqnarray}
&&\x_{ij} = a_{i,\bk} a_{j,-\bk} + a_{j,\bk} a_{i,-\bk}, \\ 
&&\x^{ij} = a_{i, \bk}^\dagger a_{j,-\bk}^\dagger + a_{j, \bk}^\dagger a_{i,-\bk}^\dagger, \\
&&\x_l^k = a_{k, \bk}^\dagger a_{l,\bk} + a_{l,-\bk} a_{k,-\bk}^\dagger
\end{eqnarray}
where $i,j,k,l = \{1,2\}$, with ten of them independent, they satisfy the commutation relations of the Lie algebra of the real symplectic group $Sp(4,R)$ \cite{perelomov_generalized_1986,hasebe_sp_2020}, i.e.
\begin{eqnarray} \label{lie}
  [\x_{ij}, \x_{kl}] &=& [\x^{ij}, \x^{kl}] = 0, \\ \nn
  [\x_{ij}, \x^{kl}] &=& \x_i^k \delta_j^l + \x_i^l\delta_j^k + \x_j^k\delta_i^l + \x_j^l\delta_i^k,\\ \nn
  [\x_{ij}, \x_l^k] &=& \x_{il}\delta_j^k + \x_{jl}\delta_i^k, \\ \nn
  [\x^{ij}, \x_l^k] &=& -\x^{ik}\delta_l^j - \x^{jk}\delta_l^i, \\ \nn
  [\x_i^j, \x_l^k] &=& \x_l^j\delta_i^k - \x_i^k\delta_l^j.
\end{eqnarray}
And the Casimir operator of this Lie algebra is given by \cite{hasebe_sp_2020}
\begin{eqnarray}
  \mathcal{C} = \x^{ij}\x_{ij} + \x_{ij}\x^{ij} - 2\x_j^i\x_i^j,
\end{eqnarray}
which commutes with all the Lie algebra generators. We can write the Hamiltonian $\hat{H}_\bk$ in terms of these generators as
\begin{eqnarray}
  \hat{H}_\bk = \xi(\bk)_i^{\ j} \x_j^i + \frac{1}{2} \eta_{ij}\x_{ij} + \frac{1}{2} \eta_{ij}^* \x^{ij}
\end{eqnarray}
where we have adopted the Einstein summation convention and $i$ is the row index, $j$ the column index for the matrices $\xi(\bk),\ \eta$. Hence, the time evolution operator $e^{-i \hat{H}_\bk t}$ gives a unitary representation of the non-compact real symplectic group $Sp(4,R)$.

Similar to the one-component BEC \cite{zhai_ultracold_2021}, the ground state of the two-component BEC Hamiltonian can be given by
\begin{eqnarray}
|G\rangle\equiv\mathcal{N} e^{\sum_{\bk\neq\mathbf{0}}\sum_{ij} Z_{\bk,ij} a_{i,\bk}^\dagger a_{j,-\bk}^\dagger} |G_0\rangle
\end{eqnarray}
where $i,j = 1,2$, and $|G_0\rangle$ is defined as $|G_0\rangle = e^{\sum_{i=1}^2\sqrt{N_i}a_{i,\bk = \mathbf{0}}^\dagger} |0\rangle$, with $|0\rangle$ the vacuum, and $Z_\bk$ is a $2\times 2$ matrix, and $\mathcal{N}$ is normalization factor.
From this equation, we can define the coherent state as
\begin{eqnarray}
  |Z_\bk\rangle &=& \mathcal{N_\bk} e^{\sum_{ij} Z_{\bk,ij} a_{i,\bk}^\dagger a_{j,-\bk}^\dagger} \va \nn\\
  &=& \mathcal{N_\bk} e^{-\frac{1}{2} Z_{\bk,ij}\mathcal{X}^{ij}} \va
\end{eqnarray}
where $\va$ is the vacuum annihilated by the operators $a_{i,\pm\bk} $, i.e. $\va = |0\rangle_{1,\bk}|0\rangle_{2,\bk}|0\rangle_{1,-\bk}|0\rangle_{2,-\bk}$, with $a_{i,\bk}|0\rangle_{i,\bk} = a_{i,-\bk}|0\rangle_{i,-\bk} = 0$, and $\mathcal{N_\bk} = \det(I - \z \z^\dagger)^{\frac{1}{2}}$ is the normalization factor (see Appendix \ref{appendix_overlap}). Here, we have adopted the Einstein summation convention. Then the ground state can be written as 
\begin{eqnarray}
  |G\rangle = |G_0\rangle \bigotimes_{\bk\neq\mathbf{0}} |\z\rangle,
\end{eqnarray}
where $\bigotimes$ denotes tensor product. 

For $|G\rangle$ to be the ground state, $|\z\rangle$ should be annihilated by the quasiparticle annihilation operators $\hat{\alpha}_{i,\bk} |\z\rangle = \sum_{j} (v_{\bk,ij}-(u_\bk \z)_{ij}) a_{j,-\bk}^\dagger |\z\rangle \equiv 0$. This equation is satisfied if $Z_\bk = u_\bk^{-1}v_\bk$ with matrices $u_\bk,v_\bk$ determined by the Hamiltonian. According to Eq.(\ref{bogo}), we have $(u_\bk^{-1}v_\bk)^T = u_\bk^{-1}v_\bk$, which means that the matrix $Z_\bk$ is symmetric. Meanwhile, we also have $I - \z^\dagger \z = u_\bk^{*-1} (u_\bk^{*-1})^\dagger $, indicating that $1 - \z^\dagger \z$ is positive definite \bibnote[Note]{For a Hermitian matrix $M$, it is positive definite if and only if it can be decomposed as a product $M = A^\dagger A$, and A is invertible. Note that $I - Z^\dagger Z$ is Hermitian.}, which can be denoted as $I - \z^\dagger \z > 0$. Hence, each coherent state $|\z\rangle$ corresponding to a point in the space defined as
\begin{eqnarray} \label{csp}
\mathcal{B} = \{Z : Z^T = Z, I - Z^\dagger Z > 0\},\nn
\end{eqnarray}
which is parameterized by 3 complex number $Z_{11}$, $Z_{22}$, $Z_{12}$ and is 6-dimension in real parameters. In mathematical literature, this space belongs to one type of the so-called Cartan classical domains, and is isometric to the quotient space $Sp(4,R)/U(2)$ \cite{perelomov_generalized_1986,coquereaux_conformal_1990}, where $U(2)$ is the unitary group of degree two. In the following, we will call the space $\mathcal{B}$ as coherent state parameter (CSP) manifold.

\section{Group action and quantum dynamics}
\label{sec:dyn}
We are interested in the quantum dynamics where initially the system is in the full condensate state $|G_0\rangle$, then it evolves according to the quenched Bogoliubov Hamiltonian $\hat{H}_\mathrm{eff}$. Since this Hamiltonian is decoupled for different $\bk, -\bk$ pairs, this allows us to study the time evolution $\hat{U}_\bk (t) \va \equiv e^{-i\hat{H}_\bk t} \va$. And in the following, for simplicity, we will drop the $\bk$ subscript.

Because the generators of $\mathfrak{sp}(4,R)$ are highly non-commuting operators, the expansion of exponential will be quite complicated. To proceed, it is very helpful to decompose the time evolution operator into a normal ordered form as \cite{perelomov_generalized_1986}
\begin{eqnarray} \label{decom}
e^{-i \hat{H} t} = e^{-\frac{1}{2} Z_{t,ij} \x^{ij}} e^{\zeta(t)_k^{\ l} \x_l^k} e^{-\frac{1}{2}\nu(t)^{ij}\x_{ij}},
\end{eqnarray}
where Einstein summation convention has been adopted. And $Z_t, \zeta(t), \nu(t)$ are $2\times 2$ matrices varying with time $t$. This decomposition is the generalization of the normal order decomposition for $SU(2)$ or $SU(1,1)$ \cite{puri_mathematical_2001}. With this decomposition, we have
\begin{eqnarray}
e^{-i\hat{H} t} |\mathbf{0}\rangle = \mathcal{N} e^{-\frac{1}{2} Z_{t, ij} \x^{ij}} |\mathbf{0}\rangle \equiv |Z_t\rangle.\nn
\end{eqnarray}
Next we will seek an expression for $Z_t$ in terms of the elements of $\h$, and prove that $Z_t$ is actually a point in the CSP manifold for any $t$.

By the virtue of the representation theory, such a decomposition should also hold true for the group element corresponding to the time evolution operator. Hence, we can convert the decomposition of the time evolution operator to the decomposition of the group element of $Sp(4, R)$.

In order to find the group element corresponding the time evolution operator, we need to rewrite the Lie algebra generators in the matrix form. Note that the generators can also be expressed as
\begin{eqnarray}
&&\x_{ij}=\hp^\dagger \beta_{ij} \hp,\ \x^{ij}=\hp^\dagger \beta^{ij} \hp,\ \x_l^k=\hp^\dagger \beta_l^k \hp \nn\\
&&(\beta_{ij})_{ab} = \delta_{a,2+i}\delta_{b,j} +\delta_{a,2+j}\delta_{b,i}, \nn\\
&&(\beta^{ij})_{ab}= \delta_{a,i}\delta_{b,2+j}+\delta_{a,j}\delta_{b,2+i}, \nn\\
&&(\beta_l^k)_{ab} = \delta_{a,k}\delta_{b,l}+\delta_{2+k,b}\delta_{2+l,a}.\nn
\end{eqnarray}
Also it is easy to verify that for any two $4\times 4$ matrices $A$ and $B$, we have
\begin{eqnarray}
[\hp^\dagger A \hp, \hp^\dagger B \hp] = \hp^\dagger \kappa [\kappa A, \kappa B] \hp,
\end{eqnarray}
where $\kappa$ is a diagonal matrix given by $\kappa = \mathrm{diag}(1,1,-1,-1)$. Therefore, we can introduce the following matrices
\begin{eqnarray}\label{matrix_generators}
Y_{ij} = \kappa \beta_{ij},\ Y^{ij} = \kappa \beta^{ij},\ Y_i^j = \kappa \beta_i^j
\end{eqnarray}
which satisfies the commutation relations of the $\mathfrak{sp}(4,R)$ Lie algebra of Eq.(\ref{lie}). Actually, the way we choose this set of matrix generators is similar to that we can choose $\{\sigma_z, \sigma^+, \sigma^-\}$ as the generators of $\mathfrak{su}(2)$ Lie algebra. As a result, we find that the symplectic matrix corresponding to the time evolution operator is $e^{-i t(\xi_i^{\ j} Y_j^i + \frac{1}{2} \eta_{ij}Y_{ij} + \frac{1}{2} \eta_{ij}^* Y^{ij})} = e^{-i \kappa \h t}$, which belongs to the $Sp(4,R)$ group. Therefore, the decomposition in the matrix form corresponding to Eq.(\ref{decom}) is
\begin{eqnarray} \label{sp_decom}
  e^{-i \kappa \h t} = e^{-\frac{1}{2} Z_{t,ij} Y^{ij}} e^{\zeta(t)_k^{\ l} Y_l^k} e^{-\frac{1}{2}\nu(t)^{ij}Y_{ij}}.
\end{eqnarray}
Since $e^{-i \kappa \h t} \in Sp(4,R)$, it can be expressed as
\begin{eqnarray} \label{sp_mx}
e^{-i \kappa \h t} =\left( \begin{array}{cc} \cu  & \cv \\ \cv^* & \cu^* \end{array} \right),
\end{eqnarray}
where $\cu , \cv$ are $2\times 2$ complex matrices satisfying $\cu \cu^\dagger - \cv\cv^\dagger = I,\ \cu \cv^T = \cv\cu ^T$ \cite{perelomov_generalized_1986,hasebe_sp_2020}. This constraint is equivalent to \cite{perelomov_generalized_1986}
\begin{eqnarray} \label{symplectic}
  &&\cu^\dagger \cu - \cv^T\cv^* = I,\nn\\
  &&\cu^\dagger \cv = \cv^T\cu^*.
\end{eqnarray}
On the other hand, through direct calculation, we find that the matrices $e^{-\frac{1}{2}Z_{t,ij}Y^{ij}},\ e^{-\frac{1}{2}\nu(t)^{ij}Y_{ij}}$ actually have upper and lower blocked-triangular form
\begin{eqnarray}
&& e^{-\frac{1}{2}Z_{t,ij}Y^{ij}} = \left( \begin{array}{cc} I & Z_t \\ 0 & I \end{array} \right),\nn\\
&& e^{-\frac{1}{2}\nu(t)^{ij}Y_{ij}} = \left( \begin{array}{cc} I & 0 \\ \nu(t) & I \end{array} \right),
\end{eqnarray}
where $I$ is $2\times 2$ identity matrix.
Meanwhile, the matrix $e^{\zeta(t)_k^{\ l} Y_l^k}$ will give a blocked-diagonal matrix as
\begin{eqnarray}
e^{\zeta(t)_k^{\ l} Y_l^k}=\left( \begin{array}{cc} (O_t^{-1})^T & 0 \\ 0 & O_t \end{array} \right),
\end{eqnarray}
where $O_t \in U(2)$. Collect all the above results, the decomposition of Eq.(\ref{sp_decom}) can be rewritten as
\begin{eqnarray}
  && \left( \begin{array}{cc} \cu & \cv \\ \cv^* & \cu^* \end{array} \right) = \nn\\ 
  && \left( \begin{array}{cc} I & Z_t \\ 0 & I \end{array} \right)
  \left( \begin{array}{cc} (O_t^{-1})^T & 0 \\ 0 & O_t \end{array} \right)
  \left( \begin{array}{cc} I & 0 \\ \nu(t) & I \end{array} \right).
\end{eqnarray}
Here the 2 by 2 matrices $Z_t$, $\nu(t)$ and $O_t$ are still unknown. By using Eq.(\ref{symplectic}), it can be directly verified that the above equation has the following solution \cite{rowe_vector_1985}
\begin{eqnarray}
Z_t &=& \cv(\cu^*)^{-1}, \\ 
\nu(t) &=& (\cu^*)^{-1}\cv^*,\nn\\ 
O_t &=& \cu^*. \nn
\end{eqnarray}
Thus, the decomposition of Eq.(\ref{sp_decom}) is satisfied. Right now, We have found the expression of $Z_t$ in terms of elements of the Hamiltonian $\h$. Next, we need to show that $Z_t$ is a point in the CSP manifold for any $t$. This can be shown by using the Eq.(\ref{symplectic}), i.e. $Z_t^T = (\cu^\dagger)^{-1}\cv^T = \cv(\cu^*)^{-1} \equiv Z_t$. Meanwhile, according to Eq.(\ref{symplectic}), we also have 
\begin{eqnarray} \label{zz}
  I - Z_t^\dagger Z_t = (\cu^*\cu^T)^{-1},
\end{eqnarray}
which implies that $I - Z_t^\dagger Z_t$ is positive definite. Hence, $Z_t$ is in the CSP manifold for any $t$.

Using the same method, we can study the time evolution of a generic coherent state $|Z_0\rangle$ with $Z_0$ in the CSP manifold. To this end, we also seek a normal ordered decomposition similar to the form of Eq.(\ref{decom}) for the operator $e^{-i \hat{H} t} e^{-\frac{1}{2} Z_{0,ij}\x^{ij}}$. By the virtue of representation theory, such a decomposition can be converted to a decomposition in the matrix level, i.e.
\begin{eqnarray}
&& \left( \begin{array}{cc} \cu & \cv \\ \cv^* & \cu^* \end{array} \right) \left( \begin{array}{cc} I & Z_0 \\ 0 & I \end{array} \right) = \nn \\
&& \left( \begin{array}{cc} I & Z_t' \\ 0 & I \end{array} \right) \left( \begin{array}{cc} (O_t'^{-1})^T & 0 \\ 0 & O_t' \end{array} \right)
\left( \begin{array}{cc} I & 0 \\ \nu'(t) & I \end{array} \right).
\end{eqnarray}
And the solution is given by \cite{rowe_vector_1985}
\begin{eqnarray}
Z_t' &=& (\cu Z_0 + \cv)(\cv^* Z_0 + \cu^*)^{-1},\\
O_t' &=& \cv^* Z_0 + \cu^*,\nn\\
\nu'(t) &=& (\cv^* Z_0 + \cu^*)^{-1}\cv^*.\nn
\end{eqnarray}
Hence it is easy to see that $e^{-i \hat{H} t} |Z_0\rangle = |Z_t'\rangle$. Therefore, the time evolution will map a coherent state to another coherent state. This also means that we can have the explicit form for the wavefunction of the system in any time of the evolution.

In summary, we have converted the problem of calculating the time evolution of a generic state $|Z_0\rangle$ with $Z_0$ in the CSP manifold into the calculation of the exponential of a matrix, i.e. $e^{-i \kappa \h t}$, which can be easily computed through numerical method. More importantly, this method also works for the case where the Hamiltonian is time dependent. In this case, the time evolution operator $\hat{U}(t)$ is defined as $\hat{U}(t) = \mathcal{T}e^{-i\int \hat{H} (t)dt}$. And the group element corresponding to the time evolution operator is 
\begin{eqnarray} \label{evo_mx}
U(t)=\mathcal{T}e^{-i\int_0^t \kappa \h(\tau)d\tau}
\equiv\lim_{N\to \infty}\prod_{n=1}^N e^{-i \kappa \h(\frac{n}{N}t) \cdot \frac{t}{N}}.
\end{eqnarray} 
Here $\mathcal{T}$ means the time ordering operator. Numerically, this can be approximated by taking $N$ as a large finite number. In this case, $U(t)$ is still a symplectic matrix, and it has the form Eq.(\ref{sp_mx}). Hence, the above method can work for the case where the Hamiltonian is time dependent.

\begin{figure}
  \subfigure[]{
    \includegraphics[width = 0.22\textwidth]{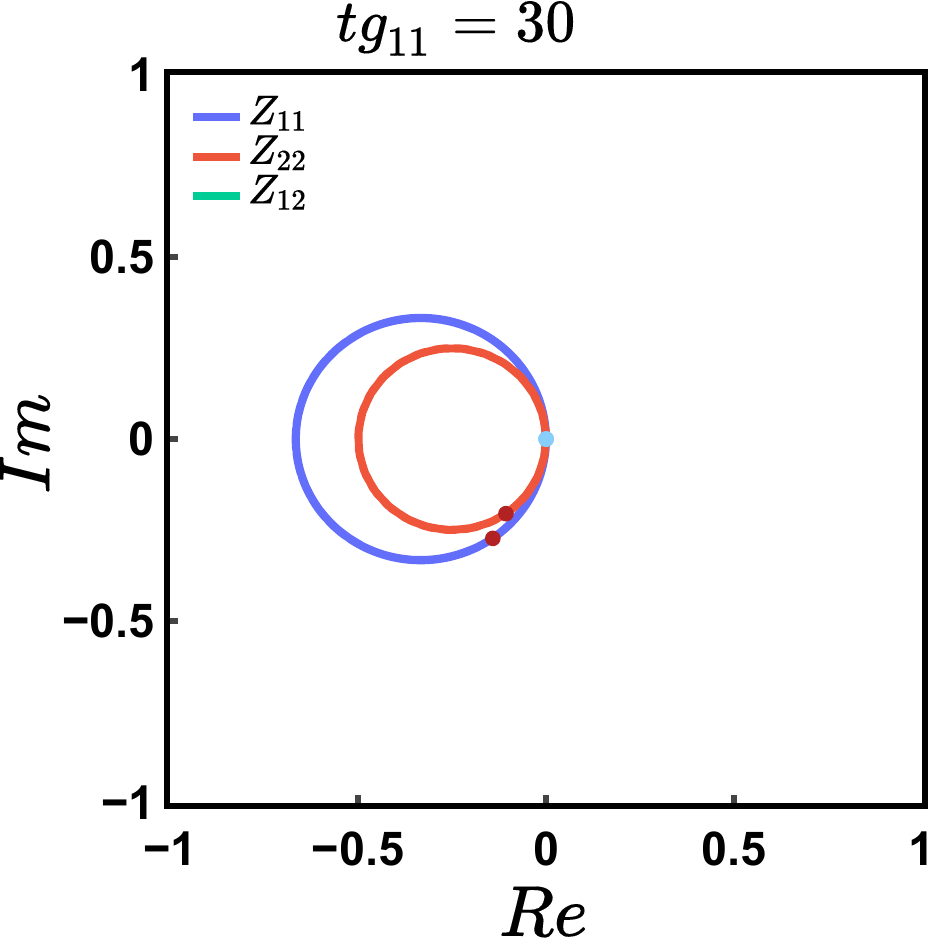}
  }
  \subfigure[]{
    \includegraphics[width = 0.22\textwidth]{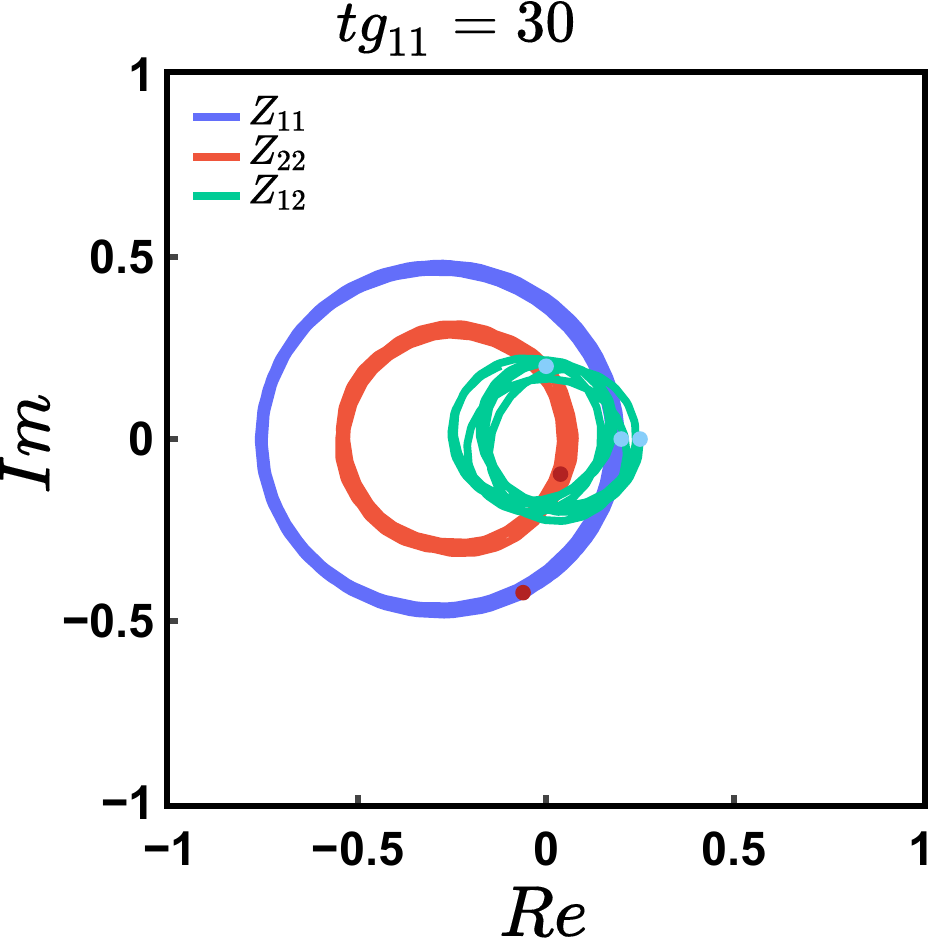}
  }
  \subfigure[]{
    \includegraphics[width = 0.22\textwidth]{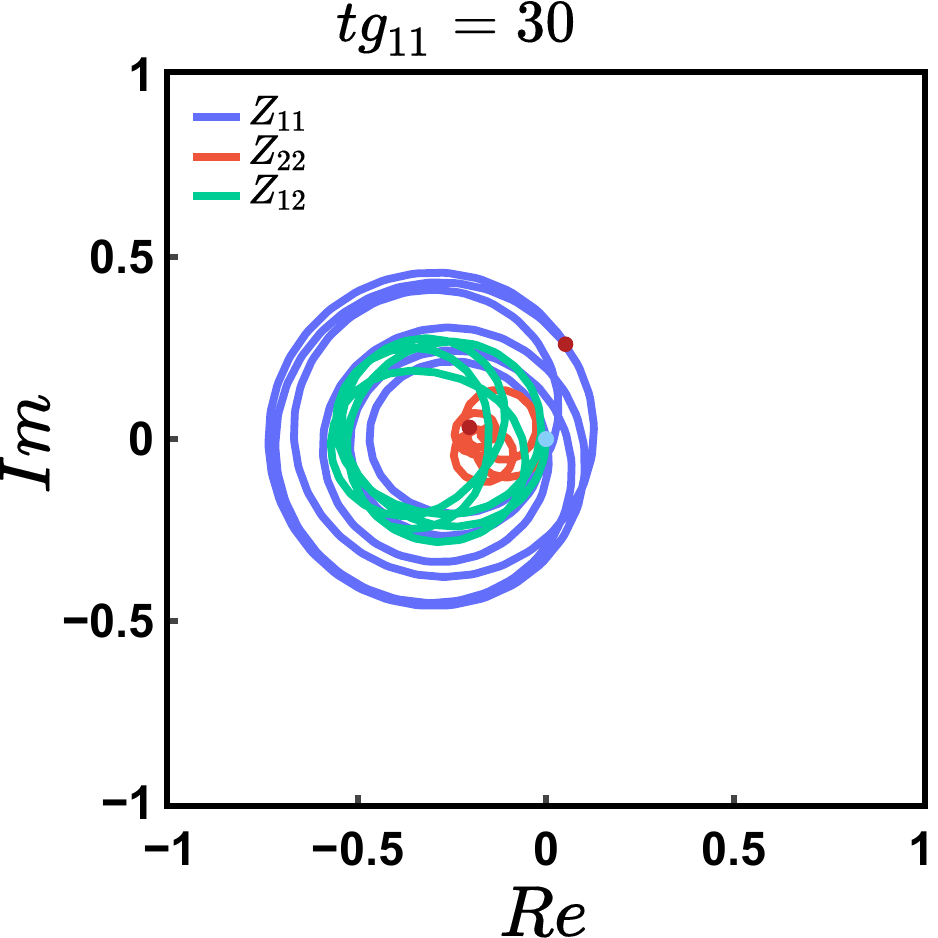}
  }
  \subfigure[]{
    \includegraphics[width = 0.22\textwidth]{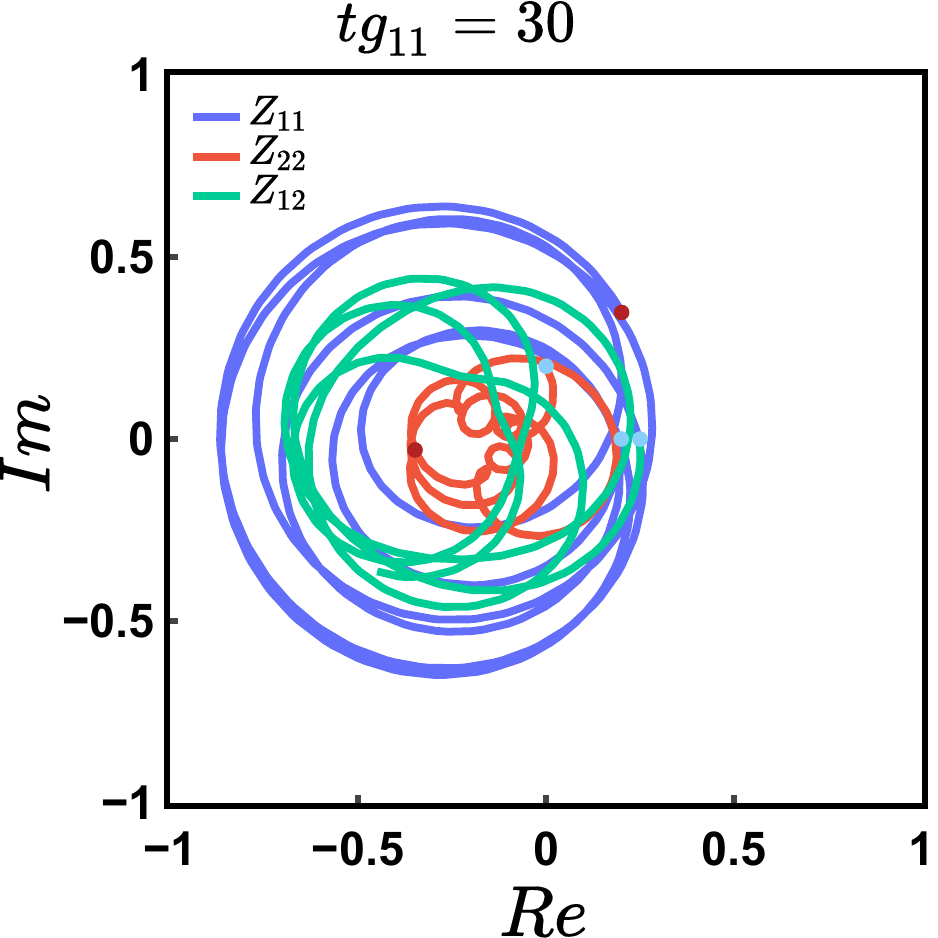}
  }
  \caption{The time evolution trajectory of complex parameter $Z_{11}$, $Z_{22}$, $Z_{12}$ on the complex plane. The light blue dot are starting point and red dot are ending point for each curve. (a)-(b) System evolves according to a decoupled Hamiltonian $\mathcal{H}_1$, starting from (a) vacuum, and (b) non-vacuum $|Z_0\rangle$ where $Z_0$ is given in the main text. (c)-(d) System evolves according to a coupled Hamiltonian $\mathcal{H}_2$ with 2 real eigenenergies, starting from (c) vacuum, and (d) non-vacuum $|Z_0\rangle$.}
  \label{dynamics}
\end{figure}

In Fig. \ref{dynamics}, we show the numerical results of trajectories of $Z_t$ in complex plane evolving under a quenched Hamiltonian with real eigenenergies. In Fig. \ref{dynamics} (a), the system starts from the vacuum $|\mathbf{0}\rangle$, and evolving according to the Hamiltonian $\mathcal{H}_1$ in which the two species of bosons are totally decoupled. The parameters of $\mathcal{H}_1$ are given by 
\begin{eqnarray}
  \xi = \left( \begin{array}{cc} 1 & 0 \\ 0 & \frac{1}{2} \end{array} \right) g_{11},\ \eta = \left( \begin{array}{cc} \frac{2}{3} & 0 \\ 0 & \frac{1}{4} \end{array} \right) g_{11}.
\end{eqnarray} 
The two real eigenenergies of $\mathcal{H}_1$ are $\varepsilon_1 \simeq 0.43 g_{11} ,\ \varepsilon_2 \simeq 0.75 g_{11}$. In this case, we can see that the off-diagonal terms are zero, and the diagonal terms $Z_{11}, Z_{22}$ form circles in the complex plane, and the evolution is periodic for the two decouple parts with different periods generally. In Fig. \ref{dynamics} (b), the system starts from a non-vacuum state $|Z_0\rangle,\ Z_0 = \left(\begin{array}{cc} 0.2 & 0.25 \\ 0.25 & 0.2i \end{array} \right)$, and evolves under the same decoupled Hamiltonian $\mathcal{H}_1$. In this case, the evolution is no longer periodic and $Z_{12}$ becomes non-zero.

In Fig. \ref{dynamics} (c-d), the systems evolve under the Hamiltonian $\mathcal{H}_2$ with two species of bosons coupled. The Hamiltonian $\mathcal{H}_2$ is given by
\begin{eqnarray}
  \xi = \left( \begin{array}{cc} 1 & \frac{1}{3} \\ \frac{1}{3} & \frac{1}{2} \end{array} \right) g_{11},\ \eta = \left( \begin{array}{cc} \frac{2}{3} & \frac{1}{2} \\ \frac{1}{2} & \frac{1}{4} \end{array} \right) g_{11},
\end{eqnarray}
whose two real eigenenergies are $\varepsilon_1 \simeq 0.28 g_{11},\ \varepsilon_2 \simeq 0.62 g_{11}$. Again, the trajectory in panel (c) starts from the vacuum $|\mathbf{0}\rangle$, the one in panel (d) starts from non-vacuum state $|Z_0\rangle$. We can see in both cases, the evolution is not periodic anymore and is also quite complicated, and the trajectory is very similar to the behavior of two coupled classical harmonic oscillators with incommensurate frequencies. The trajectory is still confined inside a finite volume of the CSP manifold. 

\section{The eigenvalues of the effective Hamiltonian}\label{eigen}
For the one-component BEC case, the behavior of the time evolution in the Poincare disk is determined by the eigenenergy of the quenched Hamiltonian, which also determines the stability of the system \cite{lyu_geometrizing_2020}. When the eigenenergy is real, the trajectory is confined in the Poincare disk, and this one-component BEC is stable; when the eigenenergy is imaginary, the trajectory approaches to the boundary of the Poincare disk, and this one-component BEC is unstable. For the two-component case, the Hamiltonian is more complicated resulting in a more complicated structure in the eigenenergies. Similar to the one-component case, the eigenenergies also determine the stability of this two-component BEC system. In this section, we will give the explicit form for the eigenenergies of the Hamiltonian. And in next section we will reveal how these eigenenergies determine the behavior of the trajectory in the CSP manifold. 

The eigenenergies of the quenched Hamiltonian are given by the solution of the equation $\det(\kappa\h - \lambda I) = 0$. According to the Cayley-Hamilton theorem, we have
\begin{widetext}
  \begin{eqnarray}
  \det(\kappa\h - \lambda I) &=& \lambda^4 - \lambda^3 \mathrm{Tr} (\kappa\h) + \frac{\lambda^2}{2}[(\mathrm{Tr} (\kappa\h))^2 - \mathrm{Tr}((\kappa\h)^2)] \nn\\
  &&  + \frac{\lambda}{6} [(\mathrm{Tr} (\kappa\h))^3 - 3 \mathrm{Tr}(\kappa\h) \mathrm{Tr}((\kappa\h)^2) + 2 \mathrm{Tr}((\kappa\h)^3)] + \det(\kappa\h).
  \end{eqnarray}
\end{widetext}
Since $\mathrm{Tr}(\kappa\h) = 0$, the above equation can be simplified to 
\begin{eqnarray}
  \det(\kappa\h - \lambda I) = \lambda^4 - \frac{1}{2} \mathrm{Tr}((\kappa\h)^2) \lambda^2 + \det(\kappa\h).
\end{eqnarray}
Hence, the eigenvalues of the matrix $\kappa \h$ can be given by $\{\lambda_+,\lambda_-,-\lambda_+,-\lambda_-\}$, where
\begin{eqnarray}
  \lambda_\pm = \frac{1}{2}\sqrt{\mathrm{Tr}((\kappa\h)^2) \pm \sqrt{\mathrm{Tr}((\kappa\h)^2)^2 - 16\det(\kappa\h)}}. \nn
\end{eqnarray}
From the above expression, we can see that $\lambda_+^2 + \lambda_-^2 = \frac{1}{2}\mathrm{Tr}((\kappa\h)^2)$. However, we also notice that $\mathrm{Tr}((\kappa\h)^2) = \tr(\xi(\bk)^2 - \eta \eta^*) + \tr(\xi^*(\bk)^2 - \eta^* \eta)$, which is always real, as $\xi(\bk)$ and $\eta$ are defined in Eq.(\ref{ham}). This gives us the following possibilities for $\lambda_+,\lambda_-$ (1) both are real; (2) one real one pure imaginary; (3) both are pure imaginary; (4) both are complex with non-zero real and imaginary parts, satisfying $\mathrm{Im}(\lambda_+^2) + \mathrm{Im}(\lambda_-^2) = 0$. Since pure imaginary eigenvalue and complex eigenvalue both have non-zero imaginary part, we will call them as complex. Hence, the above four possibilities can be simplified in to the following there cases (a) both are real; (b) one real, one complex; (c) both are complex. Since there are four eigenvalues for the matrix $\kappa \h$, for these there cases list above, we will choose the eigenenergies $\varepsilon_1, \varepsilon_2$ of the Hamiltonian in the following way, (a) $\varepsilon_1 = \lambda_+,\ \varepsilon_2 = \lambda_-$; (b) $\varepsilon_1$ is complex with $\im \varepsilon_1 > 0$, $\varepsilon_2 > 0$ is real; (c) both $\varepsilon_1,\ \varepsilon_2$ are complex, and $\im \varepsilon_1 \ge \im \varepsilon_2 > 0$. 

Hence, the eigenvalues of the matrix $e^{-i \kappa \h t}$ are $\{e^{-i\varepsilon_1 t}, e^{-i\varepsilon_2 t}, e^{i\varepsilon_1 t}, e^{i\varepsilon_1 t}\}$. The above three cases become (a) $|e^{-i\varepsilon_1 t}| = |e^{-i\varepsilon_2 t}| = 1$; (b) $|e^{-i\varepsilon_1 t}| > 1,\ |e^{-i\varepsilon_2 t}| = 1$; (c) $|e^{-i\varepsilon_1 t}| \ge |e^{-i\varepsilon_2 t}| > 1$. Actually this result is generally true for any symplectic matrix \cite{rudolph_differential_2013}. Hence, for a general time evolution matrix $U(t)$ defined in Eq.(\ref{evo_mx}), we still have the three cases listed above.

We shall emphasize that the eigenenergies also tell us about the stability of this two-component BEC system. As we shall see in Sec.(\ref{sec:obs}), by calculating the expectation value of the number operator, the system is stable only when both two eigenenergies are real, otherwise the system is unstable.

\section{The geometry of the CSP manifold}
\label{sec:geo}

For the one-component BEC, the parameter space is just a 2d Poincare disk which is easy to visualize. However, for two-component BEC case, the parameter space is six-dimensional, which is more difficult to visualize. In this section, we will introduce a polar coordinate for the CSP manifold, and show that the CSP manifold can be represented by three complex numbers, each of which is within a 2d unit disk. And with this polar coordinate, we can naturally define the boundary of this CSP manifold. Furthermore, based on this coordinate, we can show the behavior of the trajectory for all the three cases of eigenenergies from the previous section.
% \subsection{Polar Coordinate and Boundary} \label{polar_coor}

According to Autonne-Takagi factorization \cite{horn_matrix_2012}, every complex symmetric matrix can be decomposed as $u\Lambda u^T$, where $u$ is unitary matrix, and $\Lambda$ is a real diagonal with non-negative entries. This gives us the polar coordinate of the complex symmetric matrix \cite{hua_harmonic_1963}. Hence, in our case, any $Z$ in the CSP manifold can be decomposed as $Z = u \Lambda u^T$, where $u\in U(2)$ is a $2\times 2$ unitary matrix, and $\Lambda$ is a diagonal matrix $\Lambda = \mathrm{diag}(r_1, r_2)$ with $r_1 \geq r_2$. Note that every element in the group of $U(2)$ can be expressed as 
\begin{eqnarray}
u = e^{i\delta} e^{-i \frac{\varphi}{2} \sigma_z} e^{-i \frac{\theta}{2} \sigma_y} e^{-i \frac{\gamma}{2} \sigma_z}. 
\end{eqnarray}
We can absorb the phase factor $e^{i\delta} e^{-i \frac{\gamma}{2} \sigma_z}$ in $u$ to $\Lambda$, and write our new coordinate as
\begin{eqnarray}
Z = u(\varphi, \theta) \left( \begin{array}{cc} r_1 e^{i \tau_1} & 0 \\ 0 & r_2 e^{i \tau_2} \end{array} \right) u(\varphi,\theta)^T,
\label{polar}
\end{eqnarray}
where $u(\varphi,\theta) = e^{-i \frac{\varphi}{2} \sigma_z} e^{-i \frac{\theta}{2} \sigma_y}$,  and $\tau_1, \tau_2, \varphi \in [-\pi, \pi],\ \theta \in [0,\pi]$, and $\sigma_{y,z}$ are Pauli matrices. In the following, we will still call this coordinate as polar coordinate. In this coordinate, the eigenvalues of $I - Z^\dagger Z$ are $\{1 - r_1^2, 1-r_2^2\}$. Hence, the condition $I - Z^\dagger Z > 0$ is equivalent to $1 - r_1^2 > 0$, and $1-r_2^2 > 0$. And the boundary of the CSP manifold is naturally given by $\{Z(r_1,r_2,\tau_1,\tau_2,\theta,\varphi):r_1 = 1\}$, which is five-dimensional. To visualize the trajectories, we define $\lambda_1 = r_1 e^{i \tau_1}, \lambda_2 = r_2 e^{i \tau_2}, \lambda_3 = \frac{\theta}{\pi} e^{i \varphi}$, which are within the unit disk in the complex plane.

With the polar coordinate, we can study the behavior of $Z_t$ in this polar coordinate in the long-time limit, where $|Z_t\rangle = \hat{U}(t) |\mathbf{0}\rangle $. More specifically, we want to know under what condition that the coordinates $r_1(t),\ r_2(t)$ of $Z_t$ will approach to $1$, i.e. the boundary, when $t \to \infty$. We notice that $Z_t^\dagger Z_t$ is a Hermitian matrix, and its eigenvalues are $r_1(t)^2, \ r_2(t)^2$. On the other hand, according to Eq.(\ref{zz}), we have $\cu^*\cu^T = (I - Z_t^\dagger Z_t)^{-1}$. Hence, the eigenvalues of $\cu^*\cu^T$ are $\frac{1}{1-r_1(t)^2},\ \frac{1}{1-r_2(t)^2}$. Since the time evolution matrix $U_\bk(t)$ is a symplectic matrix, and can be diagonalized as $M^{-1} D M$, where $D = \mathrm{diag}(e^{-i \varepsilon_1 t}, e^{-i \varepsilon_2 t}, e^{i \varepsilon_1 t}, e^{i \varepsilon_2 t})$. As an example, we will study the case when $\varepsilon_1$ is complex and $\varepsilon_2$ is real which gives $|e^{-i\varepsilon_1 t}| > 0$  and $|e^{-i\varepsilon_2 t}| = 1$. As a result, $e^{\im\varepsilon_1 t}\to \infty$ as $t \to \infty$, and other values in $D$ are still finite as $t \to \infty$. Hence, when $t\to \infty$, we can take the approximation $D \simeq \mathrm{diag}(e^{-i \varepsilon_1 t}, 0, 0, 0)$. As a result, the matrix $\cu$ can be given by $\cu \simeq P(t)$, where $P(t)$ is a $2\times 2$ matrix, and its elements are given by 
\begin{eqnarray} \label{p_mx}
  P_{ij}(t) = e^{-i \varepsilon_1 t} M_{i1}^{-1} M_{1j}.
\end{eqnarray} 
Meanwhile, the eigenvalues of $P(t)P^\dagger(t)$ is $\{e^{2\im \varepsilon_1 t},\ 0\}$. This results in that when $t\to \infty$, one eigenvalue of $\cu\cu^\dagger$ will scale as $\sim e^{2\im\varepsilon_1 t}$, hence, approaches to infinity. Therefore, $r_1(t)$ will approach to $1$. For the case when the Hamiltonian has two complex eigenenergies, the same proof can still apply. As a result, we still have $r_1(t) \to 1$ as $t \to \infty$. However, it is still unclear about the behavior of $r_2(t)$ as $t \to \infty$ for both cases. We will address the behavior of $r_2(t)$ numerically. Furthermore, from above argument, we can also see that when the Hamiltonian has two real eigenenergies, both $r_1(t) < \infty$ and $r_2(t) < \infty$ as $t\to \infty$. Hence, in this case, the trajectory will always confine in the CSP manifold.

\begin{figure}
  \subfigure[]{
    \includegraphics[width = 0.22\textwidth]{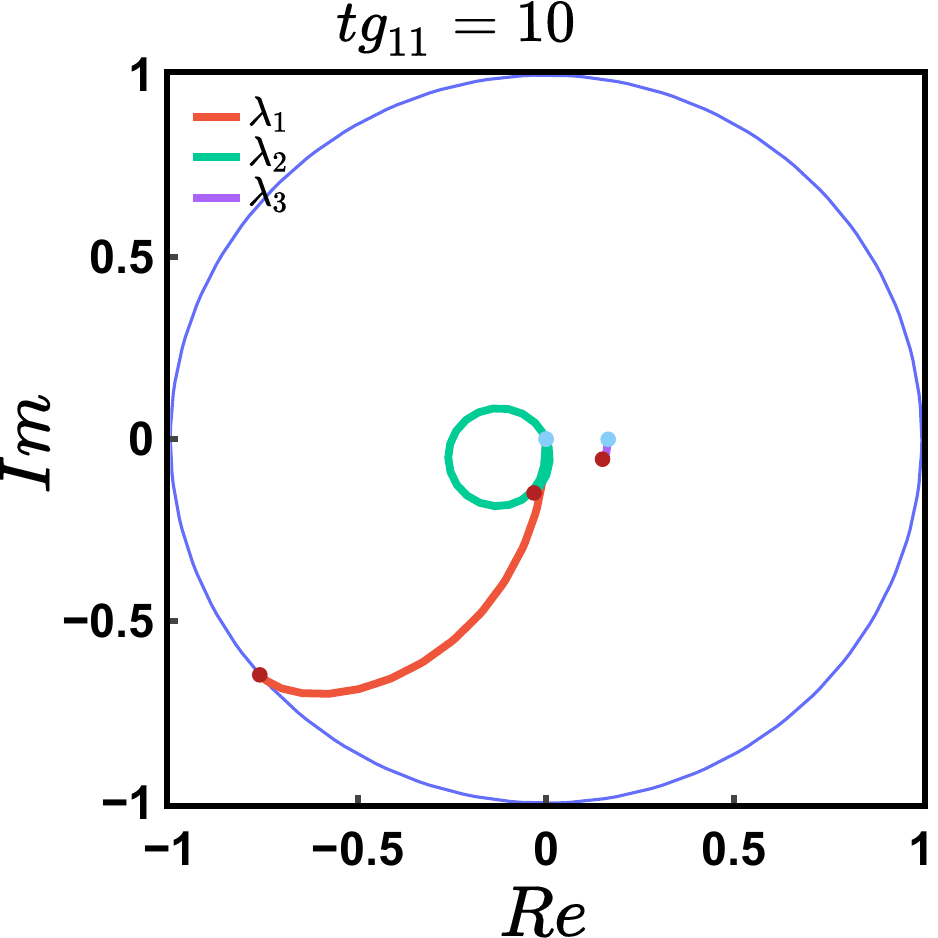}
  }
  \subfigure[]{
    \includegraphics[width = 0.22\textwidth]{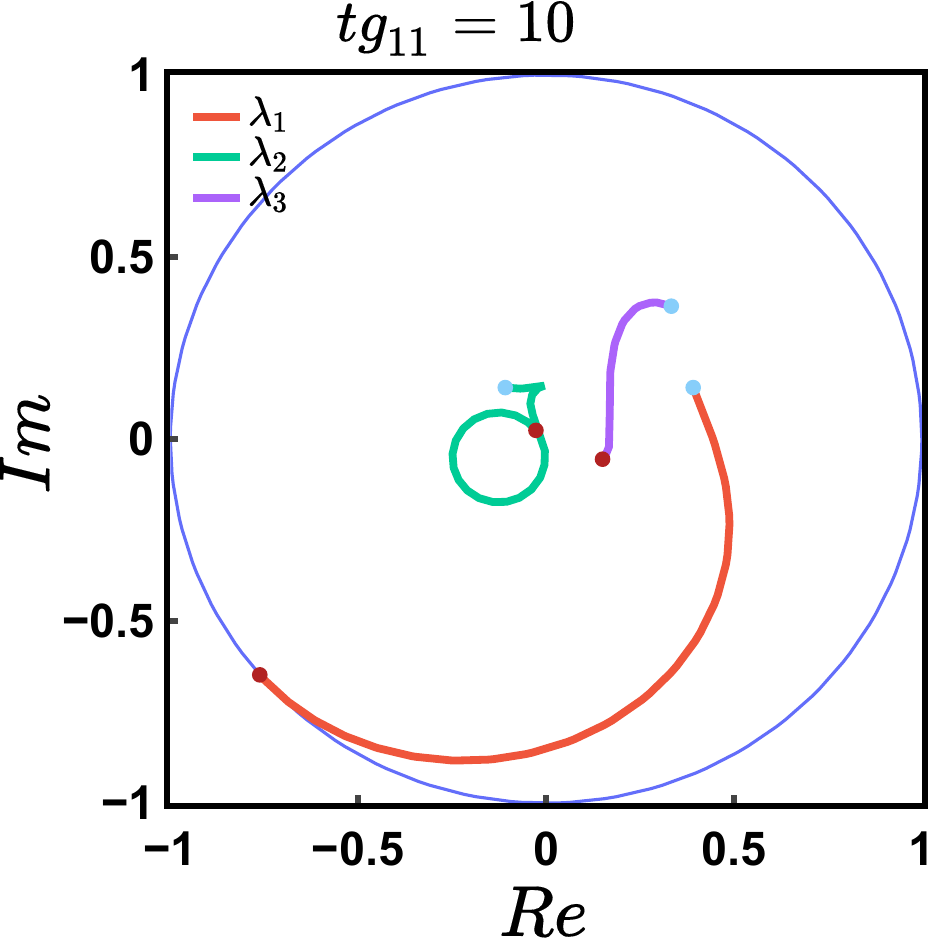}
  }
  \subfigure[]{
    \includegraphics[width = 0.22\textwidth]{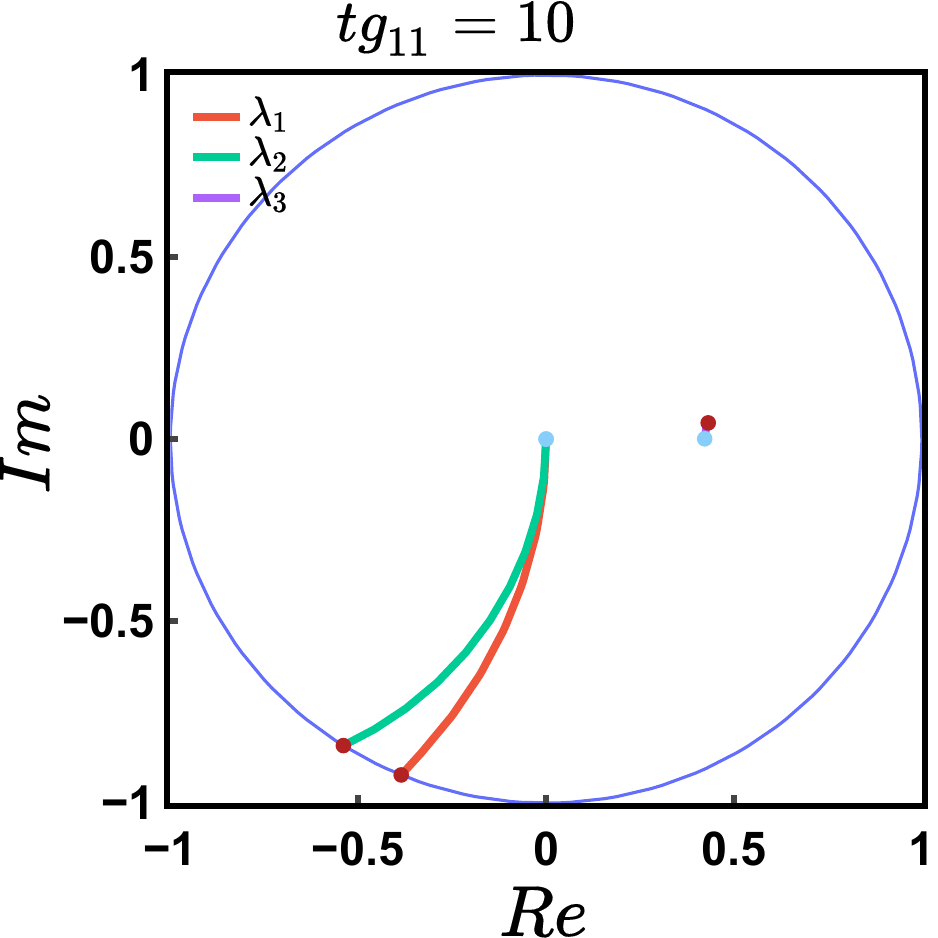}
  }
  \subfigure[]{
    \includegraphics[width = 0.22\textwidth]{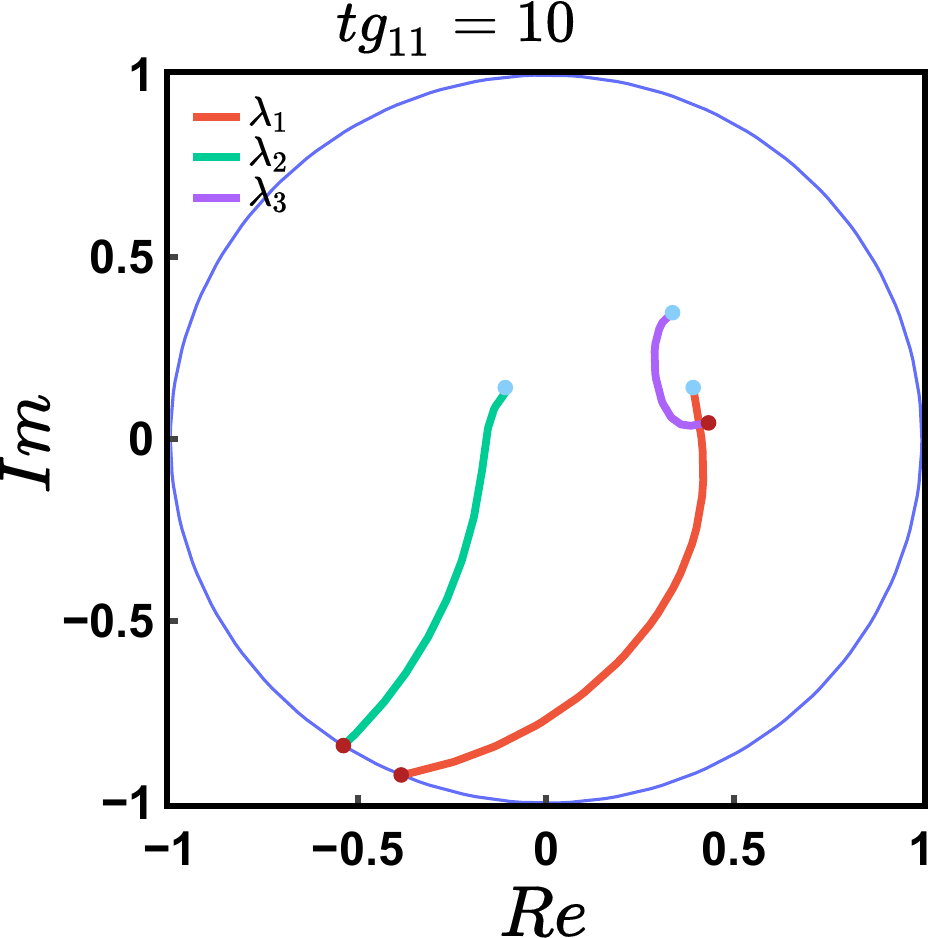}
  }
  \caption{Trajectories of quantum dynamics in the CSP manifold. The light blue dot are starting point and red dot are ending point for each curve. (a)-(b) System evolves according to a quenched Hamiltonian $\mathcal{H}_3$ with one complex eigenenergy, starting from (a) vacuum, and (b) non-vacuum $|Z_0\rangle$. (c)-(d) System evolves according to a quenched Hamiltonian $\mathcal{H}_4$ with 2 complex eigenenergies, starting from (a) vacuum, and (b) non-vacuum $|Z_0\rangle$.}
  \label{imag_dynamics}
\end{figure}

In Fig.(\ref{imag_dynamics}), we show the trajectories in the polar coordinate system we defined above. In panel (a) and (b), the time-evolution is under a quenched Hamiltonian $\mathcal{H}_3$ with the following parameters
\begin{eqnarray}
  \xi = \left( \begin{array}{cc} 1 & \frac{1}{3} \\ \frac{1}{3} & \frac{1}{2} \end{array} \right) g_{11},\ \eta = \left( \begin{array}{cc} 2 & \frac{1}{2} \\ \frac{1}{2} & \frac{1}{4} \end{array} \right) g_{11},
\end{eqnarray}
and eigenenergies $\varepsilon_1 \simeq 1.80 i g_{11},\ \varepsilon_2 \simeq 0.37 g_{11}$. The initial state is the vacuum in (a) and non-vacuum in (b). In both cases, only coordinate $\lambda_1$ approaches to the boundary, which confirms our above argument.

In panel (c) and (d), the time-evolution is under a coupled Hamiltonian $\mathcal{H}_4$ with the parameters
\begin{eqnarray}
  \xi = \left( \begin{array}{cc} 1 & \frac{1}{3} \\ \frac{1}{3} & \frac{1}{2} \end{array} \right) g_{11},\ \eta = \left( \begin{array}{cc} 2 & 1 \\ 1 & \frac{3}{2} \end{array} \right) g_{11},
\end{eqnarray}
and eigenenergies $\varepsilon_2 \simeq 2.53 i g_{11},\ \varepsilon_1 \simeq 0.6 i g_{11}$. Again, the initial state is the vacuum in (c) and non-vacuum in (d). One can see  for the Hamiltonian with two complex eigenenergies, both $\lambda_1$ and $\lambda_2$ approach to the boundary of CSP manifold. In summary, the number of coordinates approaching to the boundary equals to the number of complex eigenenergies.

\section{Dynamics of observables}
\label{sec:obs}

In this section, we will study the time evolution of the physical observable $\hat{K}$ which can be written as the linear combination of quadratic terms of the creation and annihilation operators, for example the number operator $\hat{n}_{i,\bk} = a_{i,\bk}^\dagger a_{i,\bk}, \ i = 1,2$. In the following, we will illustrate our method using the number operator. The time evolution of the number operator is $\hat{n}_{i,\bk}(t) = a_{i,\bk}^\dagger (t) a_{i,\bk}(t)$, where $a_{i,\bk}(t) = \hat{U}_\bk^\dagger (t) a_{i,\bk} \hat{U}_\bk (t)$ and $\hat{U}_\bk (t)$ is the time evolution operator. Actually, $a_{i,\bk}(t)$ is given by $a_{i,\bk}(t) = \sum_{j=1}^2 \mathcal{U}_{t,ij}a_{j,\bk} + \mathcal{V}_{t,ij} a_{j,\bk}^\dagger$ \cite{perelomov_generalized_1986}, where $\cu$ and $\cv$ is given by Eq.(\ref{sp_mx}). As a result, the time evolution of four-component operator $\hp_{\bk}(t) =  (a_{1,\bk} (t), a_{2,\bk} (t), a_{1,-\bk}^\dagger (t), a_{2,-\bk}^\dagger (t))^T$ is given by $\hp_{\bk}(t) = U_\bk(t) \hp_\bk$, where $U_\bk$ is the time evolution matrix. Hence, if we write the number operator as the form $\hat{n}_{i,\bk} = \hp_{\bk}^\dagger N_i \hp_\bk$, where $N_i$ is a $4\times 4$ matrix, given by 
\begin{eqnarray}
  N_i = \left(\begin{array}{cc} \frac{1}{2}(I-(-1)^i \sigma_z) & \mathbf{0} \\ \mathbf{0} & \mathbf{0} \end{array} \right),
\end{eqnarray}
where $I$ is $2\times 2$ identity matrix, and $\sigma_z$ is Pauli matrix. Then the time evolution of the number operator is $\hat{n}_{i,\bk} (t) = \hp_{\bk}^\dagger U_\bk^\dagger(t) N_i U_\bk (t) \hp_\bk$. As a result, all we need to do is to calculate the matrix product $U_\bk^\dagger(t) N_i U_\bk (t)$, which can be easily done by numerical method. And this method can work for any operator of the form $\hat{K} = \hp_\bk^\dagger \mathcal{K} \hp_\bk$, where $\mathcal{K}$ is a $4\times 4$ matrix.

\begin{figure}
  \subfigure[]{
    \includegraphics[width = 0.22\textwidth]{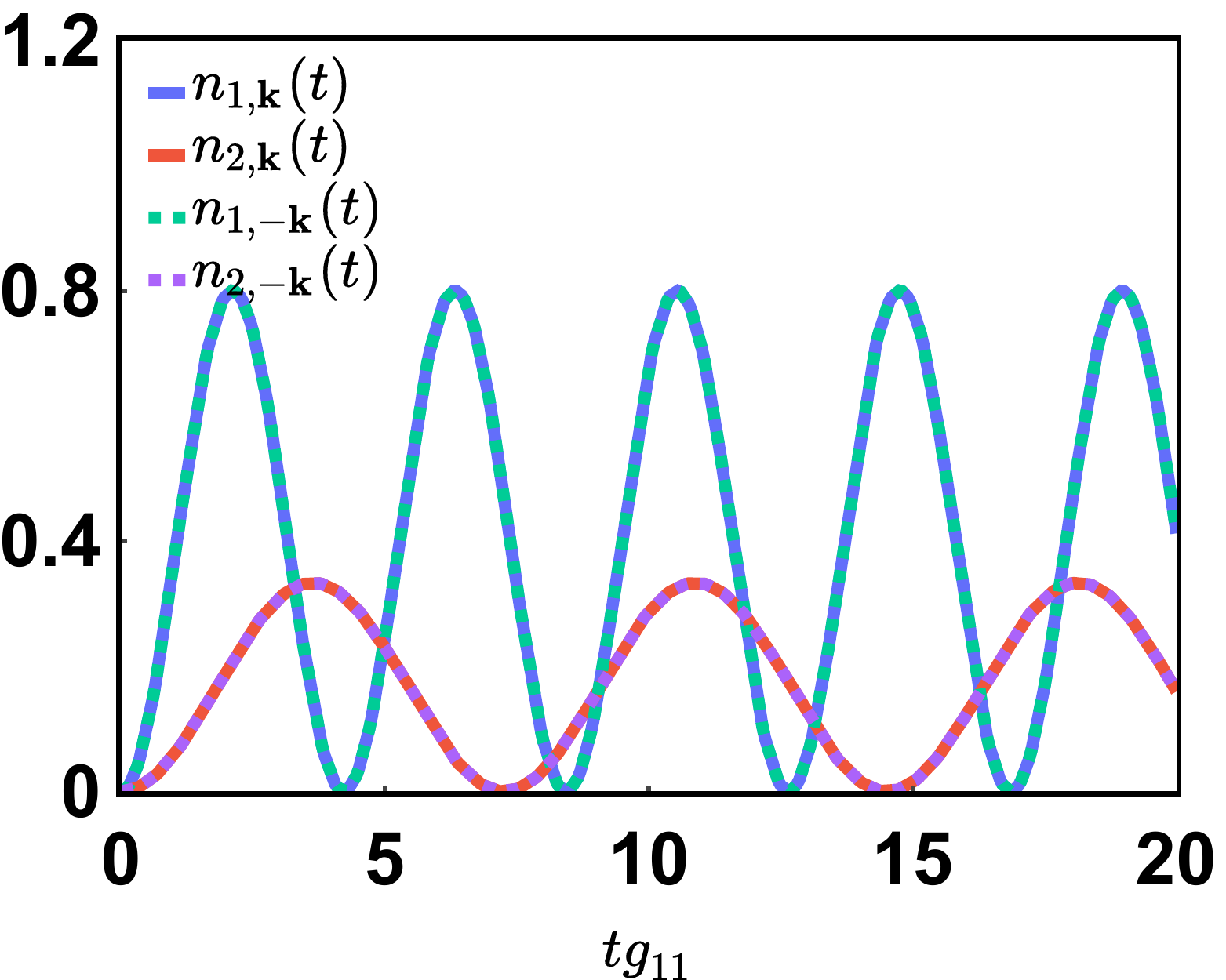}
  }
  \subfigure[]{
    \includegraphics[width = 0.22\textwidth]{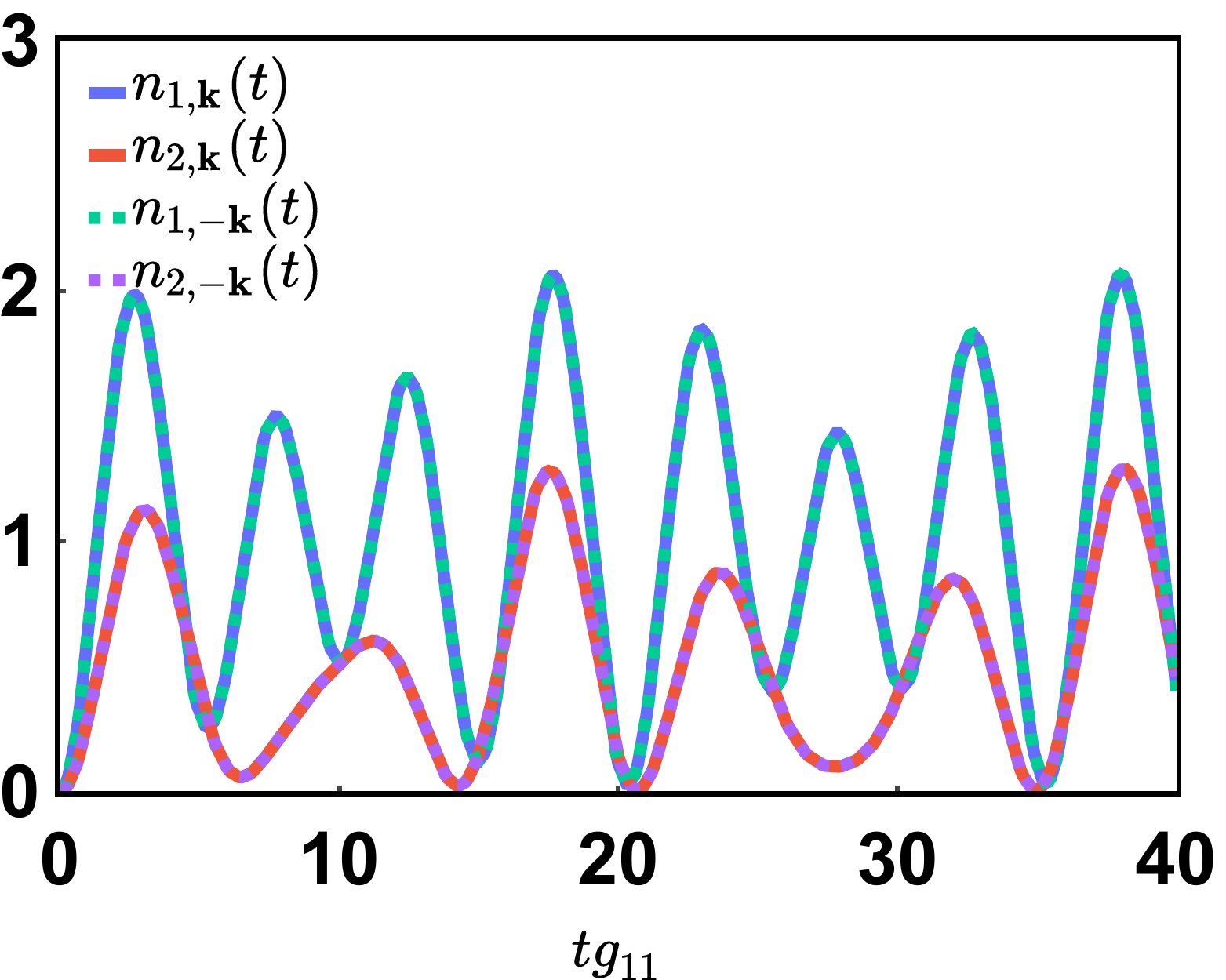}
  }
  \subfigure[]{
    \includegraphics[width = 0.22\textwidth]{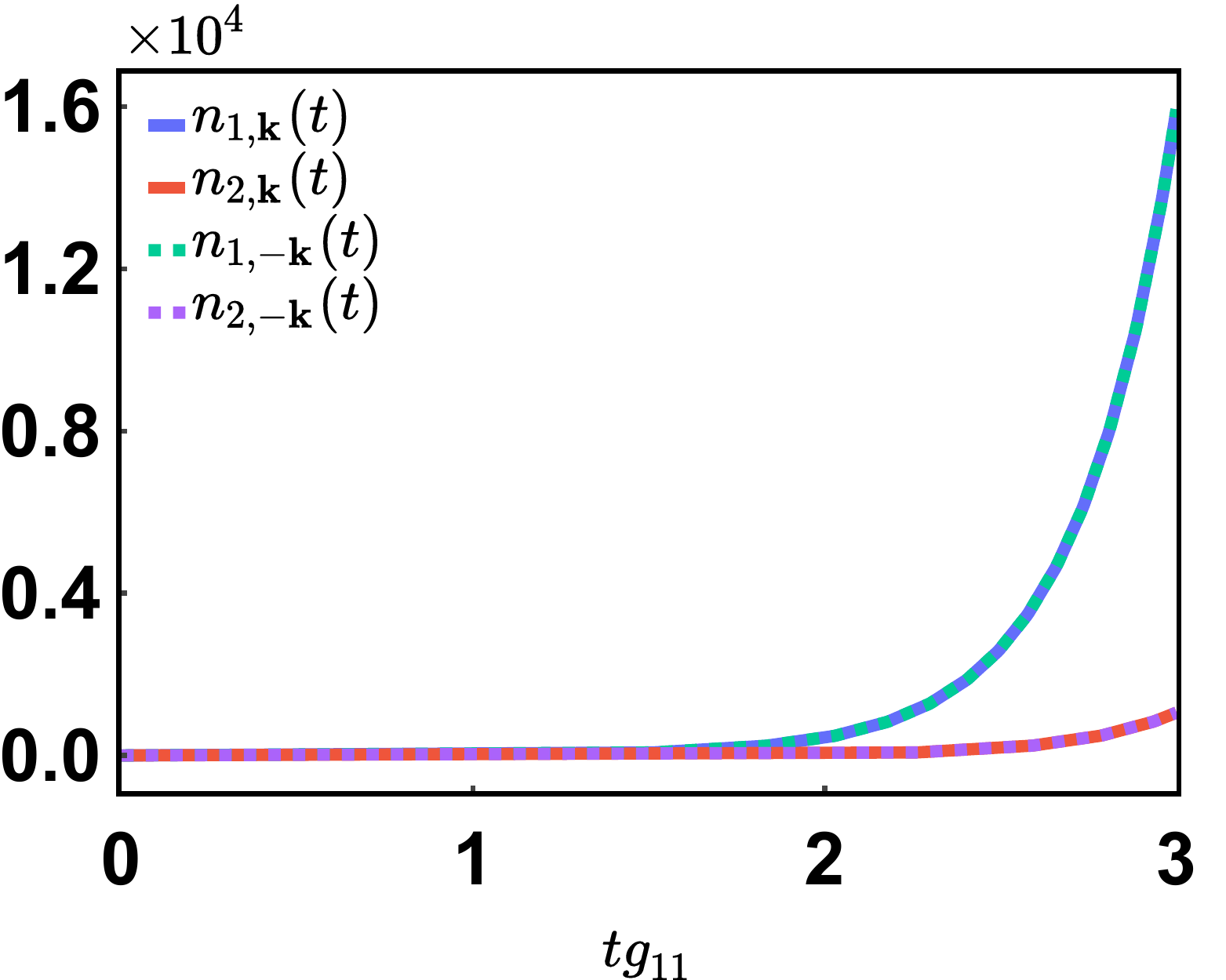}
  }
  \subfigure[]{
    \includegraphics[width = 0.22\textwidth]{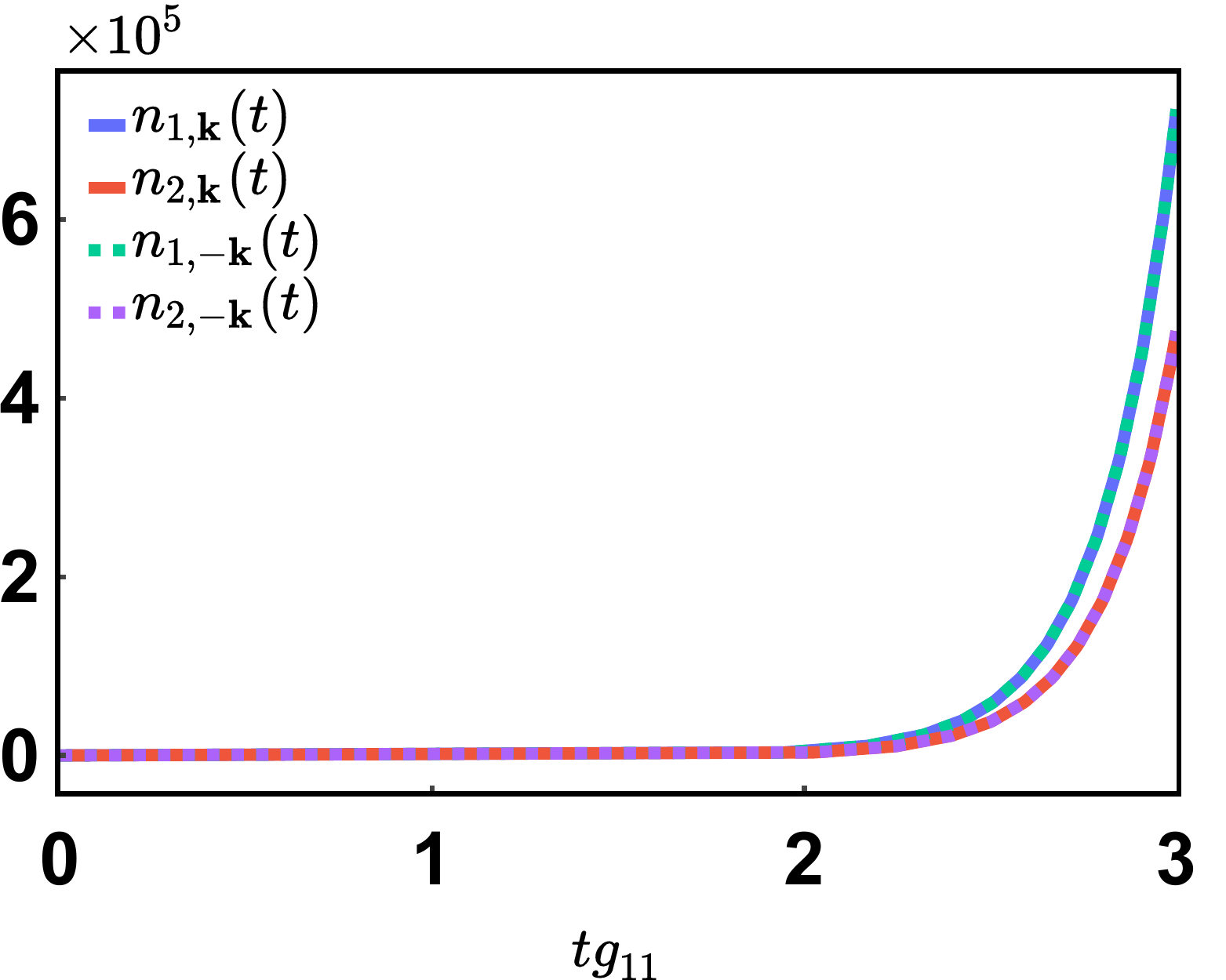}
  }
  \caption{Dynamics of number operator expectation value. In these figures, system starts from vacuum state, evolving according to (a) decoupled Hamiltonian $\mathcal{H}_1$, (b) Hamiltonian $\mathcal{H}_2$ with two real eigenenergies, (c) Hamiltonian $\mathcal{H}_3$ with one imaginary eigenenergy, (d) Hamiltonian $\mathcal{H}_4$ with two imaginary eigenenergies.}
  \label{num}
\end{figure}

Next, we will show the relation between the expectation value of number operator over the vacuum $n_{i,\bk}(t) \equiv {}_\bk\langle \mathbf{0}| \hat{n}_{i,\bk}(t) \va$ and the polar coordinate define in Sec. \ref{sec:geo}. Since only the terms ${}_\bk\langle \mathbf{0}| a_{j,-\bk} a_{j,-\bk}^\dagger \va$ contribute non-zero values to $n_{i,\bk}$, we have 
\begin{eqnarray}
  n_{i,\bk}(t) &=& -\frac{1}{2}\tr(\cv^\dagger (I-(-1)^i \sigma_z)\cv) \nn\\
  &=& -\frac{1}{2}[\tr(\cv \cv^\dagger) - (-1)^i \tr(\sigma_z \cv \cv^\dagger)]. \nn
\end{eqnarray}
By using the relation, $\cu \cu^\dagger - \cv \cv^\dagger = I$, the above equation can be rewritten as 
\begin{eqnarray}
  n_{i,\bk}(t) & = & \frac{1}{2}[\tr(\cu \cu^\dagger) + (-1)^i \tr(\sigma_z \cu \cu^\dagger)] - 1 \nn\\
  & = & \frac{1}{2}[\tr(\cu^* \cu^T) + (-1)^i \tr(\sigma_z \cu^* \cu^T)] - 1 . \nn
\end{eqnarray}
On the other hand, Eq.(\ref{zz}) results in that $\cu^* \cu^T = (I - Z_t^\dagger Z_t)^{-1}$, and using the polar coordinate defined in Eq.(\ref{polar}), we have
\begin{eqnarray}
  n_{i,\bk}(t) = \frac{1}{2}(1 + (-1)^i \cos\theta)(\frac{1}{1-r_1(t)^2} + \frac{1}{1-r_2(t)^2}) - 1. \nn
\end{eqnarray}
As we have shown in Sec. \ref{sec:geo}, if the Hamiltonian has one or two complex eigenenergies, $\frac{1}{1-r_1(t)^2} \sim e^{2 \im\varepsilon_1 t}$ as $t \to \infty$. Hence, as long as the $\theta$ is not $0$, if the Hamiltonian have either one or complex two eigenenergies, $n_{i,\bk}(t) \sim e^{2 \im\varepsilon_1 t}\to \infty$ as $t\to \infty$, for $i = 1,2$. Meanwhile, if the two species of bosons are not completely decoupled, the $\theta$ is not $0$.  As a result, in this case, the kinetic energy of bosons with the momentum $\bk$ is given by $\bk^2 n_{i,\bk}(t) /2$, which grow exponentially, indicating that the system is heated. On the other hand, when the Hamiltonian has two real eigenenergies, $n_{i,\bk}(t)$ will oscillate, which means the system is in the non-heating phase.

The above analytical result can be confirmed by numerical computation. In Fig.\ref{num}, we show our numerical results for the dynamics of the expectation value of number operator $\hat{n}_{i,\pm\bk}=a_{i,\pm\bk}^\dag a_{i,\pm\bk}$ using above method. In all these figures, the system starts from the vacuum state. In Fig.\ref{num}(a), the system evolves under the decoupled Hamiltonian $\mathcal{H}_1$. The number operator expectation value of each kind of boson oscillates according to their own period. In Fig.\ref{num}(b), the system evolves under the Hamiltonian $\mathcal{H}_2$ with two species of bosons coupled with two real eigenenergies, and the expectation values of the number operator oscillate within a range in a non-periodic way. 

In Fig.\ref{num}(c), the system evolves according to the Hamiltonian $\mathcal{H}_3$ with one complex eigenenergies, and the number operator expectation value grows exponentially for both species of bosons. But $n_{1,\pm \bk}(t)$ grow faster than $n_{2,\pm \bk}(t)$. In Fig.\ref{num}(d), the system evolves according to a Hamiltonian with two complex eigenenergy, and the expectation value of number operator grows exponentially for both species of bosons with growth rates close to each other. In addition, in all the cases, $n_{i,\bk}(t) = n_{i,-\bk}(t)$. 

Here, we shall comment on the relation between the stability of the system and the eigenenergies of the Hamiltonian. From both the analytical and numerical calculation, one can see that when the Hamilton has two real eigenenergies, the expectation values of the number operators for both species of bosons keep as a finite value, implying that in the time evolution process, the system is stable. On the other hand, when the Hamilton has one or two complex eigenenergies, the expectation values of the number operators for both species of bosons approaches to infinity as $t\to \infty$. This cannot happen in real system, which indicates that in some time this BEC will collapse and the Bogoliubov approximation is no longer valid. Thus, the system is unstable in these cases.

Meanwhile, the divergence behavior of the expectation value of the number operators are similar to the one-component BEC, where the number operator expectation value also diverges when the Hamilton has imaginary eigenenergies as discussed in Ref.\cite{lyu_geometrizing_2020}. And for the one-component case, this divergence has been shown to result in the collapse of BEC experimentally \cite{donley_dynamics_2001}.

\section{Conclusion}
\label{sec:con}
In this paper, we provide a general method to study the ground state and the quantum dynamics of the two-component Bose-Einstein condensate system. We first demonstrated that the ground state of the two-component BEC for a generic momentum $\bk$ is given by a coherent state of $Sp(4,R)$, and  can be parameterized as a point in a six dimensional manifold $Sp(4,R)/U(2)$. We then showed that the quantum dynamics of the system corresponds to trajectory in this six dimensional manifold by using the group action on this manifold. And finally, the group action on the operator also provides us a tool to calculate the expectation value of physical observables. In summary, we convert the calculation of time evolution operator to compute the exponential of  a matrix. 

Throughout this paper, we demonstrate our formalism of quantum dynamics of two-component BEC for a single generic momentum. It is also very interesting to the Floquet dynamics by periodic driven. We will leave these investigations in future works. Since our method is general one for two-component boson, it can also be applied to two-component BEC with other settings \cite{penna_twospecies_2017,richaud_quantum_2017,charalambous_control_2020}. Meanwhile, its applications are not restricted to BEC systems. Other two-component boson problem from high energy physics \cite{colas_fourmode_2022}, quantum optics or quantum information \cite{weedbrook_gaussian_2012} can also be studied by this method. Furthermore, our method can also be generalized to N-component bosonic system, the corresponding group of which is $Sp(2N,R)$.

\begin{acknowledgments}
C.Y.W. thanks Tin-Lun Ho for valuable discussions. Y. H. was supported by the Natural Science Foundation of China under Grant No. 11874272 and Science Specialty Program of Sichuan University under Grant No. 2020SCUNL210.
\end{acknowledgments}

\appendix

\begin{widetext}

\section{Calculation of the overlap between two coherent states} 
\label{appendix_overlap}

In this section, we will calculate the overlap between two coherent states $\langle Z'|Z\rangle = \mathcal{N}' \mathcal{N} \langle 0|e^{-\frac{1}{2}Z'^*_{kl}\x^{kl}}e^{-\frac{1}{2}Z_{ij}\x^{ij}}|0\rangle$. We seek to the decomposition
\begin{eqnarray} \label{overlap}
  e^{-\frac{1}{2}Z'^*_{kl}\x^{kl}}e^{-\frac{1}{2}Z_{ij}\x^{ij}} = e^{-\frac{1}{2} \mu_{ij} \x^{ij}} e^{\zeta_k^{\ l} \x_l^k} e^{-\frac{1}{2}\nu^{ij}\x_{ij}}.
\end{eqnarray}
If such a decomposition exists, then we have $\langle Z'|Z\rangle = \mathcal{N}' \mathcal{N} \langle 0|e^{\zeta_k^{\ l} \x_l^k}|0\rangle$. Furthermore, since $e^{\zeta_k^{\ l} \x_l^k} = e^{\sum_{k,l} \zeta_k^{\ l} (a_{k,\bk}^\dagger a_{l,\bk} + a_{l,-\bk}^\dagger a_{k,-\bk} + \delta_l^k )}$, we have $\langle Z'|Z\rangle = \mathcal{N}' \mathcal{N} e^{\zeta_k^{\ l} \delta_l^k} \equiv \mathcal{N}' \mathcal{N} e^{\mathrm{Tr}\zeta}$.

On the other hand, we also have the decomposition in the matrix level corresponding to Eq.(\ref{overlap}),
\begin{eqnarray}
  \left( \begin{array}{cc} I & 0 \\ -Z'^\dagger & I \end{array} \right)\left( \begin{array}{cc} I & Z \\ 0 & I \end{array} \right) = \left( \begin{array}{cc} I & \mu \\ 0 & I \end{array} \right) \left( \begin{array}{cc} (O^{-1})^T & 0 \\ 0 & O \end{array} \right) \left( \begin{array}{cc} I & 0 \\ \nu & I \end{array} \right).
\end{eqnarray}
Here, the solution can be given by $O = I - Z'^\dagger Z, \mu = Z O^{-1}, \nu = -O^{-1} Z'^\dagger$. Meanwhile, we have $(O^{-1})^T = e^{\zeta}$. Hence,
\begin{eqnarray}
  e^{\mathrm{Tr}\zeta} = \det e^{\zeta} = \det[(O^{-1})^T] = (\det O^T)^{-1}.
\end{eqnarray}
As a result, the overlap is
\begin{eqnarray}
  \langle Z'|Z\rangle = \mathcal{N}' \mathcal{N} \det(I - Z Z'^\dagger)^{-1}.
\end{eqnarray}
Taking $Z' = Z$, we have the normalization factor $\mathcal{N} = \det(I - Z Z^\dagger)^{\frac{1}{2}}$.

\end{widetext}

\bibliography{ref}

\begin{thebibliography}{35}
\bibitem [{Note()}]{Note}%
  \BibitemOpen
  Note,\ \href@noop {} {}\bibinfo {note} {For a Hermitian matrix $M$, it is
  positive definite if and only if it can be decomposed as a product $M =
  A^\dagger A$, and A is invertible. Note that $I - Z^\dagger Z$ is
  Hermitian.}\BibitemShut {Stop}%
\end{thebibliography}

\end{document}